\documentstyle[psfig]{caosp}
\begin{document}

%\pubyear{1998}
%\volume{27}
%\firstpage{449}

\htitle{Nonradial Pulsation in $\lambda$ Bootis Stars}
\hauthor{D. Bohlender {\it et al.}}

\title{The incidence of nonradial pulsation in the $\lambda$ Bootis stars}

\author{D.A. Bohlender \inst{1} \and J.-F. Gonzalez \inst{2,}
\inst{3} \and J.M. Matthews \inst{4}}
\institute{Herzberg Institute of Astrophysics, National Research Council
of Canada, 5071 W. Saanich Road, Victoria, BC, Canada V8X 4M6
\and
Centre de Recherche Astronomique de Lyon, Ecole Normale Sup\'{e}rieure de Lyon,
46 all\'{e}e d'Italie, F-069364 Lyon C\'{e}dex 07, France
\and European Southern Observatory, Casilla 19001, Santiago 19, Chile
\and Department of Physics and Astronomy, University of British Columbia,
Vancouver, BC, Canada V6T 1W5}

\date{\today}
\maketitle

\begin{abstract}
We have completed a high-resolution, high-signal-to-noise, spectroscopic
survey of the northern members of the peculiar $\lambda$ Boo stars in order
to investigate the frequency of the incidence of nonradial pulsation (NRP)
in these metal-deficient stars.  Of 18 objects observed, 9 show conclusive
evidence of NRP, which suggests that pulsation instability is a common
phenomenon in the $\lambda$ Boo class.
\end{abstract}

\section{The $\lambda$ Boo Stars}
The peculiar HgMn, Am, and magnetic Ap stars are now thought to
be reasonably well understood, but the same can not be said about
the $\lambda$ Boo stars.  These objects have Ca\,{\sc ii} K and metallic-line
types near A0, weak Mg\,{\sc ii} $\lambda$4481 lines, and H lines with cores
typical of early to late A-type stars, but often with shallow
wings.  Relative to the Balmer line cores the K and metal line
types are too early - the stars are metal weak (Gray 1988).  Quantitative
measurements suggest that CNO abundances are approximately normal
while Fe, Mg, Ca and other elements are underabundant by up to 2
dex (Venn \& Lambert 1990; St\"{u}renburg 1993).

\section{What is the Origin of the $\lambda$ Boo Stars?}
Several ideas have been put forward in the last few years to explain
the $\lambda$ Boo phenomenon.
Venn \& Lambert's (1990) observation that the abundances of elements
in 3 members of the class are similar to that of interstellar gas depleted
by the formation of grains has led to a promising model developed by
Charbonneau (1991) and Turcotte \& Charbonneau (1993) in which the
$\lambda$ Boo stars are accreting mass at a rate on the order of 
$10^{-13}M_{\odot}$ yr$^{-1}$.

If the $\lambda$ Boo stars have accreted their atmospheres,
this theory predicts that members of the class
should be accreting in the present epoch and therefore many of
them may be near the ZAMS. Since asteroseismology has the
potential to provide direct measurements of the evolutionary
state of pulsating stars, our recent discovery of
nonradial pulsation (NRP) in the $\lambda$ Boo star HD 111604 
encouraged us to complete a spectroscopic search for NRP in all
members of the class observable from the Canada-France-Hawaii
Telescope.  Our program objects are listed in Table \ref{stars}, and as an
examination of the table reveals, positive detection of NRP has
been obtained in 50\% of the 18 observed $\lambda$ Boo stars.

\begin{table}[t]
\small
\begin{center}
\caption{Program Star Data.}\label{stars}
\begin{tabular}{rlccccc}\hline\hline
HD & NRP & $\Delta t$ & $|m|$ & $P$ & Phot. Var.? & $P_{phot}$ \\
   &     &  (min)     &       & (d) &             & (min) \\ \hline
319    & No  & -- & -- &  -- & $<4.2$ mmag & -- \\
4158   & ?   &    &    &     & ?   & \\
30422  & Yes & 32 & 19 & 0.4 & Yes & 30 \\
31295  & Yes &  9 & 30 & 0.2 & $<7.4$ mmag & -- \\
38545  & No  & -- & -- &  -- & $<4.2$ mmag & -- \\
110411 & No  & -- & -- &  -- & Yes & multimode? \\
111604 & Yes & 50 & 20 & 0.7 & Yes?& \\
111786 & Yes & 39 & 17 & 0.5 & Yes & 96,43,71,46 \\
125162 & Yes & 22 & 18 & 0.3 & $<6.6$ mmag  & -- \\
142703 & No  & -- & -- &  -- & Yes & 46?,87? \\
142994 & ?   &    &    &     & Yes & 228,140,195,174 \\
183324 & Yes & 23 & 15 & 0.3 & Yes & 30 \\
192640 & Yes & 42 & 14 & 0.4 & Yes & 39,49 \\
193256 & ?   &    &    &     & $<2.6$ mmag & \\
193281 & ?   &    &    &     & $<3.4$ mmag & \\
204041 & No  & -- & -- &  -- & $<1.8$ mmag & \\
210111 & Yes & 62 & 12 & 0.5 & Yes & 51,85 \\
221756 & Yes & 49 & 17 & 0.6 & Yes & 63 \\ \hline
\hline
\end{tabular}
\end{center}
\end{table}
 
\vspace{-2mm}
\section{NRP in the $\lambda$ Boo Stars}
\begin{figure}[hbt]
\centerline{
\psfig{figure=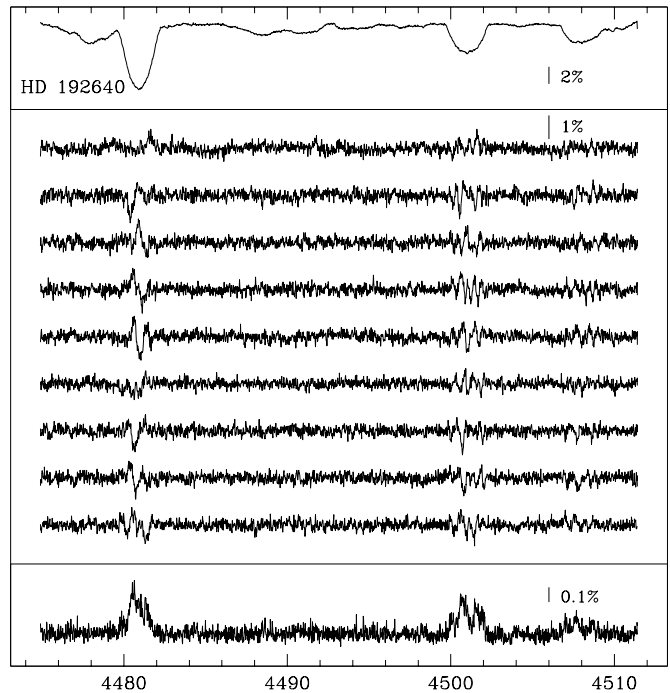,height=9.4cm}}
\caption{Observed spectra for HD 192640 in the Mg\,{\sc ii} line region.
The top panel shows the average of a sequence of spectra, the middle
panel shows the residuals of each spectrum from the mean (time increases
downwards) and the bottom shows the mean-absolute-deviation of the spectra
from the mean.  The relative intensity scale is indicated in each panel.
Individual exposures were approximately 10 min long.}
\label{fig1}
%\end{center}
\end{figure}

\vspace{-1mm}
One of the more prominent $\lambda$ Boo nonradial pulsators is shown in 
Fig. \ref{fig1}.  
Where the length of the time series for a particular star warrants,
we have attempted to estimate feature crossing times, $\Delta$t,
principal modes, $|m|$, assuming sectorial modes, and the time it
takes for one wave to travel around the star, $P$.  These values
are tabulated in Table \ref{stars}.
Also of interest is the fact that most of the nonradial pulsators
are also photometric variables. (See the column labeled ``Phot. Var.?'' in
the Table \ref{stars} for upper limits in photometric variability for
stars in which no photometric varations have been observed.)
Photometric periods are given in the last column of the table where
known.

HD 30422, HD 111604, HD 192640, and HD 210111 have the highest
amplitude NRP among our program stars.  Of these, HD 192640 (29 Cyg) has
V=4.92 which makes it very attractive as a target for future
multisite photometric and spectroscopic campaigns.  HD 31295, HD
111604, HD 183324, and HD 221756 also warrant consideration as
bright ($V<6$) pulsators.  We also would like to encourage
additional photometric observations of $\lambda$ Boo itself.

%%%%%%%%%%%%%%%%%%%%%%%%%%%%%%%%%%%%%%%%%%%%%%%%%%%%%%%%%%%%%%%%%%%%%%%%%%%%%%
%                       R E F E R E N C E S                                  %
% References should start with the \begin{thebibliography}{} command, leaving%
% the last curly brackets empty.                                             %
%%%%%%%%%%%%%%%%%%%%%%%%%%%%%%%%%%%%%%%%%%%%%%%%%%%%%%%%%%%%%%%%%%%%%%%%%%%%%%

\end{document}